\documentclass[11pt,letterpaper]{article}
\usepackage{t1enc}
\usepackage[latin1]{inputenc}
\usepackage[english]{babel}
\usepackage{graphics}
\begin{document}

\title{On the Isochronous analytic motions and the quantum spectrum}
\author{A. Raouf Chouikha
\footnote
{Universite Paris 13 LAGA, Villetaneuse 93430. 4, Cour des Quesblais 35430 Saint-Pere   
}}
\date{}
\maketitle

\bigskip

\begin{abstract} 
The problem of the characterization of all analytic potentials which give rise to isochronous oscillatory motions still open. However, there are several approaches to highlight  motions with period $T(E) \equiv T_0$ independent on the energy. 
It is proposed in this paper to give necessary and sufficient conditions for a center to be isochronous.  The proofs produced here are self contained. As corollaries we find again lot of known conditions previously established by different authors. This allows us to produce several class of isochronous analytic motions. We are also interested in the quantum spectrum. We then use the perturbation method WKB to derive an expression for the corrections to the equally spaced valid for analytic isochronous potentials.

\end{abstract}

\bigskip
\section {Introduction}
The origin of the existence problem of isochronous potentials is even older, it goes back to Huygens in 1673 [1]. The literature about isochronous potentials (or more
generally about isochronous centres) is particularly large and the interested reader is referred to [1] and references therein for more information.\\
Note however that this problem was initially posed by physicists.\ 
A tautochron curve is a curve in a vertical plane, where the time taken by a particle sliding along the curve under the uniform influence of gravity to its lowest point is independent of its point of departure.\\
The tautochron problem was solved by Huygens in the case where gravity alone acts. He proved geometrically in his Horologium oscillatorium (1673) that the curve was a cycloid. This solution was later used to address the problem of the brachistochrone curve.
This cycloide defines a sinusoidal oscillatory motion of constant pulsation
This observation thus establishes a link with the isochronous problem of the potential well. It is clear that a tautochron problem is in fact an isochronous one.

 In the plane case with rational potentials it can be shown that the only isochronous potentials with a (constant) period $ T= 2\pi$ correspond either to the harmonic oscillator $G(x) = \frac{1}{2} x^2$ or to the isotonic potential of the form $G(x) = \frac{1}{2} x^2+\frac{a}{x^2}$ up to a translation [3]. Any other isochronous potentials must be necessarily asymmetric.\\

This problem aroused a great deal of interest after the work of Landau and Lifschitz who first produced a existence condition of isochronous potentials.\\
 

The semiclassical WKB method [14] is one of the most useful approximations for computing the energy eigenvalues of the Schrodinger equation. It has a wider range of applicability than standard perturbation theory which is restricted to perturbing potentials with small coupling constants. In particular, it permits to write the quantisation condition as a power series in $\hbar$. Such series are generally non convergent. The solvable potentials are those whose series can be explicitly summed.\\

Recall at first some basic facts and introduce certain notations which will be useful in the rest of this paper.\\
Consider the scalar equation with a center at the origin $0$
\begin{equation} \qquad \ddot x + g(x) = 0 \end{equation} or its planar equivalent system
\begin{equation} \qquad \dot x = y , \qquad \dot y = - g(x)\end{equation} 

where $\dot x = \frac{dx}{dt}, \ddot x = \frac{d^2x}{dt^2}$ and $g(x)=\frac{d G(x)}{dx}$ is analytic on $R$ where $G(x)$ is the potential of (1).\\
  
 Suppose system (2) admits a periodic orbit in the phase plane with energy $E$. Given $G(x)$ and the energy $E$, Let $T(E)$ denotes the minimal period of this periodic orbit. Its expression is 
\begin{equation} T(E) = 2\int_a^b \frac{dx}{\sqrt{2E-2G(x)}}.
\end{equation} 
 $T(E)$ is well defined and is an analytic function of $E$. \\  

 We suppose that the potential $G(x)$ has one minimum value which, for convenience locate at the origin $0$ and $\frac{d^2G(x)}{dx^2}(0)=1$. The turning points $a, b$ of this orbit are solutions of $G(x) = E$.\\ Then the origin $0$ is a center of (2). This center is isochronous when the period of all orbits near $0 \in R^2$ are constant ($T = \frac {2\pi}{\sqrt{g'(0)}} = 2\pi$). The corresponding potential $G(x)$ is also called isochronous.   \\

Since the potential $G(x)$ has a local minimum at $0$, then we may consider an involution $A$ by 
$$ G(A(x)) = G(x) \ and \ A(x) x < 0 $$ for all $x \in [a,b]$. So, any closed orbit is $A$-invariant and $A$ exchanges the turning points: $b=A(a)$. 

Conversely, by using this involution we may calculate for a prescribed period function $T(E)$ the distance between the turning points 
$$T(E) = 2\int_0^c [\frac{dA(a)}{dG}-\frac{d a}{dG}]\frac{dG}{\sqrt{2E-2G}}.$$
which implies 
\begin{equation} A(a) - a = b - a = \frac {2\pi}{\pi }\int_0^E \frac  {d\gamma}{\sqrt{2E - 2\gamma}}.\end{equation}

 The potential is isochronous means the period is constant ($T = 2\pi$),\ then for all orbits near zero  one has necessarily
$$b - a =\frac {1}{\pi }\int_0^E \frac {T(\gamma) d\gamma}{\sqrt{2E - 2\gamma}}= 2 \sqrt{2E}.$$\\ 

In the first section we give others conditions for the potential to be isochronous and we precise his link with the older ones. 
We then review some properties of isochronous potentials and derive new analytical expressions for some multiparameters
families of isochronous potentials. This work improves our previous paper [5].\\ The first two quantum corrections $I_2$ and $I_4$ of the WKB series for the torus quantization
for the 1D (one-dimensional) stationary Schrodinger equation have been explicitly expressed by Robnik and Romanovski in [13]. Dorignac [6] has calculated these first 2 corrections in some situations thanks to the Abel type integrals (or Riemann-Liouville fractional). We show that it is the same for the higher corrections $I_{2n}, n \geq 2$ for isochronous potentials. Thus, it allows to give higher order corrections, resulting in a convergent series whose sum is identical to the exact
spectrum.

\section{Isochronicity conditions for the potential\\ energy}
 
 \subsection{Statement of results}

The problem to determine whether the center is isochronous has attracted many researchers for long time. This problem has been recently revived due to advancement of computer algebra. New powerful algorithms have been discovered indeed. We will propose an alternative approach in order to derive more directly new isochronous potentials. More precisely, we propose to give necessary and sufficient conditions for a oscillatory motion to be isochronous. Our conditions seem
more natural since it allows us to deduce all the others well known before. Other equivalent characterizations will be considered.  More exactly, we state the following

\bigskip

{\bf Theorem A} \quad {\it  Let $g(x)$ be an analytic function and $G(x) =\int_0^x g(s) ds $ and $A$ be the analytic involution defined by $G(A(x)) = G(x)$. Suppose that for\ $x\neq 0, x g(x) > 0$.\ Then the  equation $$ \ddot x + g(x) = 0 \qquad (1)$$  has an isochronous center at $0$ if and only if the function $$\frac{d}{dx}[G(x)/g^2(x)]$$ is $A$-invariant i.e. 
$\frac{d}{dx}[G/g^2](x) = \frac{d}{dx}[G/g^2](A(x)) $
 in some neighborhood of $0$. }\\

\bigskip 

In order to study  the period function $T(E)$ depending on the energy, it is sometime convenient to study its derivatives. 
We need the following \\

{\bf Lemma 2-1} \quad {\it The
 derivative of the period function (depending on the energy) $T'(E) = \frac{dT}{dE}$ may be written}
\begin{equation} T'(E) = (\frac {1}{E})\int _a^b \frac {g^2(x) - 2 G(x) g'(x)}{g^2(x) \sqrt{2E - 2G(x)} } dx\end{equation}

 This lemma has been initially proved by  Chow and Wang [6] but the proof given in [5] is more direct.\\
 

The following result proved in [5] explicitly highlights the role and impact of involution on the behavior of the period $T \equiv T(E)$\\

\bigskip
 {\bf Lemma 2-2}\quad {\it The derivative of the period function $T(c)$ may also be written as}
$$T'(E) = (\frac {1}{E}) \int_0^{b}\frac{\frac{d}{dx}(\frac{G}{g^2}(x))-\frac{d}{dx}(\frac{G}{g^2}(A(x))}{\sqrt{2E-2G(x)}}g(x) dx.$$



\bigskip
{\it \bf Proof of Theorem A}\\
Suppose hypothesis of Theorem A \ $ \quad \frac{d}{dx}[G(x)/g^2(x)] $ \ is $A$-invariant holds, then by Lemma 2-2
$$ T'(E)  = (\frac {1}{E})\int _a^b \frac {\frac{d}{dx}(G/g^2)}{ \sqrt{2E - 2G(x)} } g(x)dx = (\frac {1}{E}) \int_0^{b}\frac{\frac{d}{dx}(\frac{G}{g^2}(x))-\frac{d}{dx}(\frac{G}{g^2}(A(x))}{\sqrt{2E-2G(x)}}g(x) dx$$
It implies obviously\ $T'(E) \equiv 0$\ and the center $0$ of equation (1) is isochronous \\
Conversely, let \ $T'(E) \equiv 0$\ then by Lemma 2-2 
$$T'(E) = (\frac {1}{E}) \int_0^{b}\frac{\frac{d}{dx}(\frac{G}{g^2}(x))-\frac{d}{dx}(\frac{G}{g^2}(A(x))}{\sqrt{2E-2G(x)}}g(x) dx = 0.$$

Since\ $b > 0$\ and the function \ $\frac{d}{dx}(\frac{G}{g^2}(x))-\frac{d}{dx}(\frac{G}{g^2}(A(x))$\ is analytic, then necessarily   \ $\frac{d}{dx}(\frac{G}{g^2}(x))=\frac{d}{dx}(\frac{G}{g^2}(A(x))$.\\
That means the function \ $\frac{d}{dx}(\frac{G}{g^2}(x))$\ is $A$-invariant.\\

We may also prove the following 

\bigskip 

{\bf Theorem B} \quad {\it Let $G(x) =\int_0^x g(s) ds $ be an analytic potential. Suppose that for\ $x\neq 0, x g(x) > 0$.\ Then the  equation $$ \ddot x + g(x) = 0 \qquad (1)$$  has an isochronous center at $0$ if and only if  
\begin{equation} x - \frac{2 G}{g} = F(G) \end{equation} 
where $F$ is an analytic function defined in some neighborhood of $0$. }\\

\bigskip  
{\bf Corollary 2-3 } \quad {\it  Under the hypotheses of Theorem B, equation (1)  has an isochronous center at $0$ if and only if  
\begin{equation}\frac{d}{dx}[G(x)/g^2(x)] = f(G) \end{equation} 
where $f$ is an analytic function defined in some neighborhood of $0$. }\\

\bigskip 
 This Corollary has been proved in [5] as a main result (Theorem B)\\

{\it  Proof of Corollary 2-3 }\\
Condition\ $ x - \frac{2 G}{g} = F(G)$ \ implies by Theorem B that  $$\frac{d}{dx}[x - \frac{2 G}{g}] = \frac{d}{dx}F(G) =1- 2\frac{g}{g}+2\frac{G g'}{g^2} = -f(G) = -1+2\frac{G g'}{g^2}$$. Conversely, let us consider $F$ the primitive of $f $ such that $F(0)=0 $. Then \\ $$ \frac{d}{dx}[G(x)/g^2(x)] = f(G(x)) \Leftrightarrow g(x)\frac{d}{dx}[G(x)/g^2(x)] = g(x)f(G(x)).$$  Integrate by parts, it yields \ $ \frac{G(x)}{g(x)}- x = F(G(x)).$ \. Thus it is equivalent to the condition \ $\frac{d}{dx}[G/g^2](x) = f(G)$. \ Thus, one gets another equivalent condition of isochronicity.\\ 

\bigskip
{\it \bf Proof of Theorem B } \quad 
Starting from the derivative  $$ T'(E)  = (\frac {1}{E})\int _a^b \frac {\frac{d}{dx}(G/g^2)}{ \sqrt{2E - 2G(x)} } g(x)dx $$ which may be expressed after integrate by parts
$$T'(E)  = (\frac {1}{E})\int _a^b \frac{1}{3}\frac {\Phi(x)}{ (\sqrt{2E - 2G(x)})^3 } g(x)dx$$
where $$\Phi (x) = \int_0^x {\frac{d}{dx}(G/g^2)} g dx = \frac{2G}{g} - x$$ Thus, $$T'(E)  = (\frac {1}{3E})\int _a^b \frac{(\frac{2G}{g} - x)}{ (\sqrt{2E - 2G(x)})^3 } g(x)dx.$$
When \ $T'(E) \equiv 0$\ therefore \ $x -\frac{2 G}{g}$\ is $A$-invariant and \ $x -\frac{2 G}{g}$\ is an analytic function dependent on $G$ by the following lemma.\\

\bigskip
 {\bf Lemma 2-4}\quad {\it Consider the function \ $\phi(G,x)=\frac{d}{dx}[G(x)/g^2(x)]$.\ Suppose in addition \ $\phi(G,x)$\ is $A$-invariant then its derivatives with respect to $G$ are all $A$-invariant for $x$ in a neighborhood of $0$ (i.e. $\frac{d^p \phi(G,A(x))}{dG^p} = \frac{d^p \phi(G,x)}{dG^p}$ for any integer $p > 0$}.

\bigskip 
 {\it Proof} \quad We will proceed by recurrence. Notice that $$\frac{d \phi(G,x)}{dG}=\frac{d}{dG}\frac{d}{dx}[G(x)/g^2(x)]=\frac{d}{dx}\frac{d}{dG}[G(x)/g^2(x)]$$ since the functions $G$ and $g$ are analytic. Moreover, one gets
$$\frac{d}{dx}\frac{d}{dG}[G(x)/g^2(x)]= \frac{1}{g(x)}\frac{d^2 }{dx^2}[\frac{G}{g^2}](x)$$
On the other hand, we have seen that \\ $G(A(x))=G(x)$\ implies \ $g(A(x)) \frac{d A(x)}{dx} = g(x).$\ Since \ $\phi(G,x)$\ is $A$-invariant, then $$\frac{d}{dx}[G/g^2](x) = \frac{d}{dx}[G/g^2](A(x))$$ implies in deriving another time $$\frac{d^2}{dx^2}[G/g^2](x) = \frac{d^2}{dx^2}[G/g^2](A(x)) \frac{d A(x)}{dx}. $$ Therefore 
{\footnotesize $$\frac{1}{g(A(x))}\frac{d^2 }{dx^2}[\frac{G}{g^2}(A(x))]= \frac{1}{g(A(x))\frac{d A(x)}{dx}}\frac{d^2 }{dx^2}[\frac{G}{g^2}(A(x)]\frac{d A(x)}{dx} = \frac{1}{g(x)} \frac{d^2}{dx^2}[G/g^2](x)$$}
That means \ $\frac{d}{dx} \phi(G,x)$\ is A-invariant.\\

Suppose now the following derivative  \ $\phi_n(G,x)=\frac{d^n \phi(G,x)}{dG^n}$ \ is $A$-invariant for any integer\ $ n\leq p.$ \
Then  $$\frac{d^{n+1} \phi(G,x)}{dG^{n+1}}=\frac{d}{dG}\frac{d^n \phi(G,x)}{dG^n}=\frac{1}{g(x)}\frac{d}{dx}\frac{d^n}{dG^n}[G(x)/g^2(x)]=\frac{1}{g(x)}\frac{d}{dx}\phi_n(G,x). $$ Since \ $\phi_n(G,A(x))=\frac{d^n \phi(G,A(x))}{dG^n}= \frac{d^n \phi(G,x)}{dG^n}=\phi_n(G,x)$\ then, deriving this expression one obtains
$$\frac{d}{dx}\phi_n(G,x) = \frac{d}{dx}\phi_n(G,A(x))\frac{d A(x)}{dx}.$$ Thus, 
{\footnotesize $$\frac{1}{g(A(x))}\frac{d}{dx}\phi_n(G,A(x))= \frac{1}{g(A(x))\frac{d A(x)}{dx}}\frac{d }{dx}\phi_n(G,A(x))\frac{d A(x)}{dx} = \frac{1}{g(x)}\frac{d}{dx}\phi_n(G,x).$$}
That means $$\phi_{n+1}(G,x) = \frac{d^{n+1} \phi(G,x)}{dG^{n+1}} = \frac{d^{n+1} \phi(G,A(x))}{dG^{n+1}}= \phi_{n+1}(G,A(x)).$$ 

Some other equivalent conditions may also be deduced, [5]

\bigskip
{\bf Corollary 2-5} \quad {\it Under the hypothesis above\ $0$ is an isochronous center of (1) if and only if $x=x(G)$ is an analytic solution of the linear ODE 
\begin{equation} 2G \frac{d^2x}{dG^2}+\frac{dx}{dG} = f(G),\end{equation}
where $f$ is an analytic function. \\ Moreover, this solution must satisfy the conditions:} $$x(0)= 0,\quad lim_{G\rightarrow 0}(\frac{x^2}{2G})=1.$$



{\bf Corollary 2-6} \quad {\it  Let $G(x) =\int_0^x g(s) ds $ be an analytic potential of the scalar equation $$ \ddot x + g(x)= 0. \qquad (1)$$ \\ Let $A(x)$ be  an analytic involution defined by: \\ $  
G(A(x)) = G(x)$ \ and \ $A(x)x < 0$.\\ Then $0$ is an isochronous center of (1) if and only if  $$\frac{d}{dx}[G(x)/g^2(x)] = \frac{4 A''(x)}{(1-A'(x))^3}. $$   Moreover, the last expression is $A$-invariant. }\\

\bigskip 

 This involution $A(x)$ may also be defined as a solution of a linear ODE. The following result is analogous to Corollary 2-5.

\bigskip 
{\bf Corollary 2-6} \quad {\it Under hypotheses of Theorem A, equation (1) admits an isochronous center at $0$ if and only if the involution $A=A(G)$ is a solution of 
\begin{equation} 2G \frac{dA}{dG}= A(G) + F(G)
\end{equation}
 $f$ is an analytic function and $F$ is its integral.\\ Moreover, this solution must satisfy the conditions:} $$A(0)= 0,\quad lim_{G\rightarrow 0}(\frac{A^2}{2G})=1$$



\section{Links with previous results }
 
\subsection{Condition of Landau and Lifschitz }

This result has several consequences and corollaries. first of all we find again the condition established by Landau and Lifschitz [12]. We point out that these authors are the first to provide an excellent starting point by deducing the constraint placed on the form of the potential by a prescribed energy dependence of the period.
 They proved the following 
\bigskip                     

 {\bf Proposition 3-1} \ [12, Chap.3]  {\it Let $G(x) =\int_0^x g(s) ds $ be an analytic potential. Suppose that for\ $x\neq 0, x g(x) > 0$.\ (1) has an isochronous center at the origin $0$ if and only if 
 $$x - A(x) = 2\sqrt{2 G(x)}$$
 for all $0<x<b$ where $A$ is the involution such that $G(A(x)=G(x)$ and $A(x) x < 0$ for $x \neq 0$.} 

\bigskip
{\it \bf Proof of Proposition 3-1} \qquad We need the following lemmata (the first one has been proved in [5]) \\

 {\bf Lemma 3-2}\quad {\it For any analytic involution $A(x)$ defined for all $x \in [a,b]$ the following expression holds }
\begin{equation}
 \frac{d}{dx}[\frac{2}{(1-A(x)^2}] = \frac{4 A''(x)}{(1-A'(x))^3}
\end{equation}


\bigskip
 {\bf Lemma 3-3}\quad {\it When the function $\frac{d}{dx}[\frac{G(x)}{g^2(x)}]$\ is $A$-invariant then the following holds}
$$\frac{d}{dx}[\frac{G(x)}{g^2(x)}] = \frac{d}{dx}[\frac{2}{(1-A(x)^2}] = \frac{4 A''(x)}{(1-A'(x))^3}$$

\bigskip 
 {\it Proof} \quad Indeed a direct calculation yields
$$\frac{d}{dx}[\frac{G}{g^2}](A(x)) = \frac{d}{dx}[\frac{G(x)}{g^2(A(x))}] =[\frac{g^2-2Gg'}{g^3}](A(x)) $$
But $$[\frac{g^2-2Gg'}{g^3}](A(x))= \frac{g^2 A'^{-2}-2 G(g' A'^{-2}-g A'' A'^{-3})}{g^3 A'^{-3}}(x)$$
$$=A'(x) (\frac{g^2-2Gg'}{g^3})(x) + (\frac{2 G A''}{g^2})(x)$$
It implies
$$(\frac{g^2-2Gg'}{g^3})(x)\ (1 - A'(x)) = (\frac{2 G A''}{g^2})(x)$$
Or equivalently
 $$\frac{\frac{d}{dx}[\frac{G(x)}{g^2(x)}]}{\frac{G(x)}{g^2(x)}} = \frac{2 A''(x)}{1 - A'(x)}$$
We then obtain by integration
$$\frac{G(x)}{g^2(x)} = \frac{K}{(1 - A'(x))^2}$$
where the constant $K$ necessarily equals $2$ since \\ $A'(0)=-1$ \ and\ $lim_(x\rightarrow 0) \frac{G(x)}{g^2(x)} = \frac{1}{2}$ .\\ 
Therefore 
$$\frac{G(x)}{g^2(x)} = \frac{2}{(1 - A'(x))^2}$$
By deriving we also obtain
$$\frac{d}{dx}[\frac{G(x)}{g^2(x)}] = \frac{4 A''(x)}{(1-A'(x))^3}$$

\bigskip 

On the other hands we may deduce the following
$$1-A'(x) = \frac{g(x)\sqrt 2}{\sqrt G(x)}$$
By integration we then obtain
\begin{equation}
x - A(x) = 2\sqrt{2G(x)}\qquad for \quad all \qquad 0<x<b.
\end{equation} 
Conversely, starting from \ $x - A(x) = 2\sqrt{2G(x)}$\ and deriving twice one gets 
$$\frac{d}{dx}[\frac{G(x)}{g^2(x)}] = \frac{4 A''(x)}{(1-A'(x))^3}$$
This implies Proposition 3-1 since $A(0)=G(0)=g(0)=0$.

\subsection{ Theorem of Kouckles and Piskounov} 

When $g(x)$ is a continuous function, $g(x)$ and $x$ having the same sign,  Koukles and Piskounov [10] produced necessary and sufficient conditions so that the center of the system (2) is isochronous.  

\bigskip
 {\bf Proposition 3-4} \ [10, Th 5] \quad {\it A set of necessary and sufficient conditions for the period of every solution of (1) near $0$ to be equal to a constant $T_0$ is:\\
 1 - $g(x)$ is continuous and positive for small positive $x$.\\
 2 - $liminf_{x\rightarrow 0} \mid\frac{g(x)}{x}\mid \neq 0.$\\
 3 - $T_0 \geq limsup_{x\rightarrow 0} \frac{2\pi}{\sqrt{g(x)/x}}.$\\
 4 - $g(-x) = - \frac{d}{dx}[\frac{T_0}{\pi}\sqrt{2x}-(\int_0^xg(u) du)^{-1}]^{-1}$\ where index $-1$ denotes an inverse function.}

\bigskip

When $g(x)$ is a analytic function, $g(x)$ and $x$ having the same sign,  Koukles and Piskounov [10] produced a necessary and sufficient condition so that the center of the system (2) is isochronous. 
 
\bigskip
 {\bf Proposition 3-5} \ [10, Th 6] \quad {\it Let $g(x)$ be a real analytic function.  Then the center $0$ of the equation $(1)\quad \ddot x+g(x) = 0$ is isochronous if and only if the inverse function $x = \Theta(z)$ \ 
 $$ [\int_0^x g(\xi) d\xi ]^{-1} = \Theta(z)$$ is of the form
 $$ \Theta(z) = \sqrt{z}+ P(z)$$ where $P$ is a real analytic function such that $P(0) =0$ . }
 
  \bigskip
	{\it \bf Proof of Proposition 3-5} \qquad \\ This result is a direct consequence of Corollary 3-3. Indeed, consider the change \ $x = \sqrt{2G} + y$\ we have proved that \ $y=P(G)$ is a unique solution of $$ 2G \frac{d^2y}{dG^2}+\frac{dy}{dG} = f(G)$$ with initial points : $y(0) =0, y'(0)= 1.$\  A resolution of the last equation yields 
$$y(G)= \sqrt{2G}\int_0^G \frac{F(\nu)}{(2\nu)^{3/2}}d \nu.$$ Conditions\ $y(0) =0, y'(0)= 1.$,\  imply this solution \ $y=P(G)$\ is necessarily analytic. Thus, a solution of (9) may be written
$$x(G)= \sqrt{2G}\ (1+ \int_0^G \frac{F(\nu)}{(2\nu)^{3/2}}d \nu)= \sqrt{2G} + H(G)$$ where \ $P(G)$\ is analytic in $G$. The proof is achieved.\\

\subsection{Urabe Theorem}

  Later, Urabe proposed some refinements of Proposition 3-5 by considering the assumption of the differentiability of $g(x)$ at $0$ and proved the following result which is most used than Propositions 3-5 because he considers less rigid hypotheses, the functions are only class $C^1$.
   
\bigskip
 
 {\bf Proposition 3-6} \ [17] \quad {\it Let \ $ g(x) $\ be a \ $C^1$ \  function defined in $V_0$ a neighborhood of $0$ verifying   $x g(x) > 0$ in $V_0 /\{0\}$. Then the system $(2)$ has an isochronous center at the origin $0$ if and only if   \  $g(x)$\ may be written 
 $$g(x) = \frac {X}{1 + h(X)}$$ 
 where \ $h(X)$\ is a \ $C^1$ \ odd function and \ $X = \sqrt{2 G(x)},\ \frac {X}{x} > 0$\ for \ $x\neq 0$.}\\

 \bigskip
{\it \bf Proof of Proposition 3-6} \qquad In the analytic case, Theorem B and Corollary 3-3 together imply that $$x = \frac{2G}{g} + F(G) = \sqrt{2G} + P(G) .$$ Thus, 
$$\frac{2G}{g} = \sqrt{2G} + \psi(G)$$ (where \ $\psi(G)$\ is a analytic function in $G$) is another necessary and sufficient condition for the potential of equation (1) to be isochronous. Simplify by  $\sqrt{2G}$ it yields  $$\frac{\sqrt{2G}}{g} = 1 + \frac{\psi(G)}{\sqrt{2G}}$$ which is equivalent to the Urabe condition 
 $$g(x) = \frac{\sqrt{2G}}{1 + \frac{\psi(G)}{\sqrt{2G}}} =  \frac {X}{1 + h(X)}$$ where $X = \sqrt{2G}$

 \bigskip

{\bf Remark 3-7} \quad As it has noticed in [5] the method of Urabe used an intermediary function $h$ that is not in general explicitly known.   Our condition are simply relations between an isochronous potential $G$ and its derivative $g$. Thus, any analytic solution \ $G=G(x)$\   of  Equation $(6)\ 2G(x) - xg(x) = g(x) F(G(x)) $ \ or equivalently Equation $(7)\ \frac{d}{dx}[G(x)/g^2(x)] = f(G(x)) $\  provides an isochronous  potential  verifying $G(0)=G'(0)=0, G''(0)=1$. We will illustrate this fact by several multiparameters families of isochronous potentials.\\ 

\subsection{Others consequences}

The next result is natural, although unexpected. Yet he will play a special role in order to perform the WKB expansion to all orders for any isochronous potential.\\

\bigskip 
{\bf Proposition 3-8} \quad {\it Let $G(x) =\int_0^x g(s) ds $ be an analytic potential defined in a neighborhood of $0$. Suppose equation $$ \ddot x + g(x) = 0 \qquad (1)$$  has an isochronous center at $0$. Let $g^{(n)}(x)$\ be the n-th derivative of the potential (with respect to $x$): \ $g^{(n)}(x) = \frac{d^n}{dx^n} G(x), \ n \geq 1$\ then these derivatives may be expressed under the form
$$g^{(n)}(x) = a_{n}(G) \sqrt{2G} + b_{n}(G), n \geq 0$$
where \ $a_n $ and $b_n$ are analytic functions  with respect to $G$}.

\bigskip 
In fact, as we will see in the sequel, the functions $a_{n}$ and $b_{n}$ are only dependent on $G_1$ the odd part of $G=G(x)$.\\
\bigskip 

{\it \bf Proof of Proposition 3-8}\qquad  By Proposition 3-4, condition\\ $x(G)  = \sqrt{2G} + P(G)$ \ with $P=P(G)$ is a non-zero analytic function  implies that equation (1) an isochronous center at $0$. Deriving with respect to $G$ one obtains $$\frac{dx}{dG} = \frac{1}{\sqrt{2G}} + P'(G) = \frac{1}{g}$$ or equivalently
$$\frac{g}{\sqrt{2G}} = \frac {1}{1+\sqrt{2G} P'(G)} = a_1(G) + \frac{b_1(G)}{\sqrt{2G}}$$ with 
$$a_1(G) = \frac {-1}{2GP'^2-1} \quad and \quad b_1(G) = \frac {2GP'}{2GP'^2-1}.$$
Notice that by hypothesis $G$ is defined defined in a neighborhood of $0$ then $2GP'^2-1 $ is necessary non zero.\\
The functions $a_1(G)$ and $b_1(G)$ are analytic since $P$ and $P'$ they are too.\\ Derive now $g'(x)$ it yields
$$g'(x) =\frac {d g}{dx} = {\frac {d}{dx}}a_1 \left( G \right) \sqrt {2}\sqrt {G}+1/2\,{
\frac {a_1 \left( G \right) {g}\sqrt {2}}{\sqrt {G}}}+{\frac {
d}{dx}}b_1 \left( G \right) 
 =$$ $$ {\frac {d}{dG}}a_1 \left( G \right) g\sqrt {2}\sqrt {G}+1/2\,{
\frac {a_1 \left( G \right) {g}\sqrt {2}}{\sqrt {G}}}+{\frac {
d}{dG}}b_1 \left( G \right) g.$$ 
$$g'(x) = \left(  \left( {\frac {d}{dG}}{ a_1}   \right) 
\sqrt {2}\sqrt {G}+1/2\,{\frac {{ a_1}  \sqrt {2}}{
\sqrt {G}}}+{\frac {d}{dG}}{ b_1}   \right)  \left( 
{ a_1}  \sqrt {2}\sqrt {G}+{ b_1}   \right) $$
where the symbol prime $'$ means $\frac{d}{dG}$ and $a_1$ or $b_1$ stands for $a_1(G)$ or $b_1(G)$.

After replacing\ $g(x) = a_1(G) \sqrt{2G} + b_1(G)$ \ one obtains
$$2\, G { a'_1}    { a_1}
  +  { a'_1}  
  \sqrt {2}\sqrt {G}{ b_1}  +  { a_1}
    ^{2}+1/2\,{\frac {{ a_1} \sqrt {2}{ b_1}  }{\sqrt {G}}}+ \ { b'_1}    { a_1}  \sqrt {2}\sqrt {G}+  { b'_1}   { b_1}  
$$

By simplifying one find the expression of \ $g'(x) =a_2(G) \sqrt{2G} + b_2(G)$ \ with $$a_2(G)=a'_1 b_1 + \frac{a_1 b_1}{2 G} + b'_1 a_1$$
$$b_2(G) = 2 G a_1 a'_1 + a_1^2 + b_1 b'_1 $$
Here $$\frac{a_1 b_1}{2 G} = \frac{P'}{(2 G P'^2-1)^2}$$ which is analytic. 
Then the functions $a_2(G)$ and $b_2(G)$ are analytically dependent on the functions $a_1(G), b_1(G)$ and their derivatives.\\

Suppose now that until order $p$ one has $$g^{(p)}(x)= a_p(G)+b_p(G)$$ where the function $a_p(G)$ and $b_p(G)$ are analytic with respect to $G$. Thank to {\it Maple} we are able to carry out the calculations. Deriving this expression with respect to $x$ it yields

$$g^{(p+1)}(x)= {\frac {d}{dx}}a_p \left( G \right) \sqrt {2}\sqrt {G}+1/2\,{
\frac {a_p \left( G \right) {g}\sqrt {2}}{\sqrt {G}}}+{\frac {
d}{dx}}b_p \left( G \right)  =$$ $$ a'_p \left( G \right) g\sqrt {2}\sqrt {G}+1/2\,{
\frac {a_p \left( G \right) {g}\sqrt {2}}{\sqrt {G}}}+b'_p \left( G \right) g.$$
$$=\left( a'_{{p}} \left( G \right)  \right)  \left( a_{{1
}}\sqrt {2}\sqrt {G}+b_{{1}}   \right) \sqrt {2}\sqrt 
{G}+1/2\,{\frac {a_{{p}} \left( G \right) \left( a_{{1}}\sqrt {2}\sqrt {G}+b_{{1}}
   \right) \sqrt {2}}{\sqrt {G}}}+$$ $$ \left( b'_{{p}} \left( G \right)  \right)  \left( a_{{1}}\sqrt {2}
\sqrt {G}+b_{{1}}   \right) $$
By simplifying one obtains
$$g^{(p+1)}(x)= 2\,Ga_{{1}}a'_p{{{ }}}+a_{{1}}a_{{p}}+\sqrt {G}a_{{1}}\sqrt {2}b'_p{{
{}}}+\sqrt {G}b_{{1}} a'_p{{{}}}\sqrt {2}+1
/2\,{\frac {a_{{p}}\sqrt {2}b_{{1}}  }{\sqrt {G}}}+b_{
{1}}  b'_p{{{}}}
$$ 
where the symbol prime $'$ means $\frac{d}{dG}$ and $a_p$ or $b_p$ stands for $a_p(G)$ or $b_p(G)$.\\
By simplifying one find the expression of \ $g^{(p+1)}(x) =a_{p+1}(G) \sqrt{2G} + b_{p+1}(G)$ \ with $$a_{p+1}(G)=a'_p b_1 + \frac{a_p b_1}{2 G} + b'_p a_1$$
$$b_{p+1}(G) = 2 G a_1 a'_p + a_1 a_p + b_1 b'_p .$$

Moreover, since $$\frac{a_p b_1}{2 G} =  \frac {a_p P'}{2GP'^2-1}$$ then the fonction \ $\frac{a_p b_1}{2 G}$\ is analytic with respect to $G$. Thus,
 $a_{p+1}(G)$ \ and \ $b_{p+1}(G)$ \ are also analytic.\\

Notice that the coefficients $a_{n}(G)$ et $b_{n}(G)$ depend analytically on $P$ and its derivatives. It should be mentioned that their complexity 
 increases rapidly with the order $n$ except obviously for the first terms. Namely, after replacing $a_1$ and $b_1$ in the expression of $a_2$ and $b_2$ and after simplifying we get 
$$a_2 =a'_1 b_1 + \frac{a_1 b_1}{2 G} + b'_1 a_1=\frac{2G P'^3 + 12G^2 P''P'^2 + 3P'+ 2G P''}{(2G P'^2-1)^3}$$
$$b_2=2 G a_1 a'_1 + a_1^2 + b_1 b'_1=-\frac{6G P'^2 + 12G^2 P''P' +1+  8G^3 P''P'^3}{(2G P'^2-1)^3}.$$
The next one is already complicated enough
$$a_3 = {a_1}'' {b_1}^2+2a'_1b'_1b_1+2{b_1}''a_1b_1+4a_1^2a'_1+2G{a'_1}^2a_1+2Ga_1^2{a_1}''+b_1^2a_1$$ $$+ \frac{4a'_1b_1^2+4b'_1 a_1b_1-a_1b_1^2}{4G}$$
$$b_3 = 4G{a_1}'' b_1a_1+4Ga'_1b'_1a_1-\frac{a_1^2b_1}{2G}+2G{b_1}''a_1^2+2a_1^2b'_1+2G{a'_1}^2b_1+b_1^2{b_1}''+{b'_1}^2b_1$$

\bigskip

{\bf Remark  1 } \quad it seems very likely that we can show the following result: {\it Let $G(x)$ be an analytic potential such that $x g(x) > 0, x\neq 0$. Suppose the nth derivative may be expressed as \ $g^{(n-1)}(x) = a_{n}(G) \sqrt{2G} + b_{n}(G),\ n \geq 0$\ where $a_n$ and $b_n$ are analytic functions, then the potential is isochronous}.\\
We need for that to solve already the differential system for a given $g^{(1)}(x) = g'(x)$       
\begin{eqnarray}
a_2 & = & a'_1 b_1 + \frac{a_1 b_1}{2 G} + b'_1 a_1\nonumber\\
b_2 & = & 2 G a_1 a'_1 + a_1^2 + b_1 b'_1
\end{eqnarray}
where the symbol prime $'$ means $\frac{d}{dG}$ and $a_p$ or $b_p$ stands for $a_p(G)$ or $b_p(G)$. Moreover the initial conditions are \
$a_1(0)=1, b_1(0)=0.$\  Which means that the class of isochronous potentials is rigid.\\ 

{\bf Remark  2 } \quad In fact, in the case of isochronism knowing  \ $a_1(G) = \frac {-1}{2GP'^2-1},\quad  b_1(G) = \frac {2GP'}{2GP'^2-1}$\  we then obtain a simple relation between them after eliminating $P'$
\begin{eqnarray}
b_1^2 = 2G (a_1^2-a_1).\end{eqnarray}
 Which means that given an analytic potential $G$ and 2 functions $a_1(G)$ and $b_1(G)$ verifying relation (13) so that $g(x) =\frac{dG}{dx}$ can be expressed \\ $g(x) = a_{1}(G) \sqrt{2G} + b_{1}(G)$ \ then $G$ is an isochronous potential.\\

\bigskip
Another result no less interesting which on the one hand is more general than the lemma 2-2 and on the other hand it will be very useful for the sequel in the WKB approximation for isochronous potentials.\\

\bigskip
{\bf Proposition 3-9}\quad {\it Let $G(x) =\int_0^x g(s) ds $ be an analytic potential and $\phi(x)$ an function defined in a neighborhood of $0$. $A$ be the analytic involution defined by $G(A(x)) = G(x)$. Then for $a < 0 < b=A(a)$ and $G(a)=G(b)= E$ the following integrals equality holds
$$\int_a^b \frac{\phi(x)}{\sqrt{E-G(x)}} g(x) dx = \int_0^b \frac{\phi(x)-\phi(A(x))}{\sqrt{E-G(x)}} g(x) dx$$
In particular, if we may expressed $\phi(x) = u(G)\sqrt{2G} + v(G)$ then }
$$\int_a^b \frac{\phi(x)}{\sqrt{E-G(x)}} g(x) dx = \int_0^b \frac{2u(G)\sqrt{2G}}{\sqrt{E-G(x)}} g(x) dx$$

\bigskip 
{\bf Proof of Proposition 3-9} \qquad It suffices to split the integral
$$\int_a^b \frac{\phi(x)}{\sqrt{E-G(x)}} g(x) dx = \int_a^0 \frac{\phi(x)}{\sqrt{E-G(x)}} g(x) dx + \int_0^b \frac{\phi(x)}{\sqrt{E-G(x)}} g(x) dx$$
Recall that $ a < 0 < b $. By definition when $x \in [a,0]$ then $A(x) \in [0,b]$ and conversely. By a change of variable $x=A(y)$ the integral becomes
{\footnotesize $$\int_a^0 \frac{\phi(x)}{\sqrt{E-G(x)}} g(x) dx = -\int_0^b \frac{\phi(A(y))}{\sqrt{E-G(y)}} g((A(y)) A'(y) dy = -\int_0^b \frac{\phi(A(y))}{\sqrt{E-G(y)}} g(y) dy$$}
since $g((A(y)) A'(y) = g(y)$. Therefore
 $$\int_a^b \frac{\phi(x)}{\sqrt{E-G(x)}} g(x) dx = \int_0^b \frac{\phi(x)}{\sqrt{E-G(x)}} g(x) dx -\int_0^b \frac{\phi(A(y))}{\sqrt{E-G(y)}} g(y) dy= .$$
On the other hand, since $\phi(x) = u(G)\sqrt{2G} + v(G)$. Then the following integral may be written 
$$\int_a^b \frac{\phi(x)}{\sqrt{E-G(x)}} g(x) dx = \int_a^b \frac{u(G)\sqrt{2G}+v(G)}{\sqrt{E-G(x)}} g(x) dx =$$ $$ \int_a^b \frac{u(G)\sqrt{2G}}{\sqrt{E-G(x)}} g(x) dx + \int_a^b \frac{v(G)}{\sqrt{E-G(x)}} g(x) dx$$
The last integral can be written
$$ \int_a^b \frac{v(G)}{\sqrt{E-G(x)}} g(x) dx = \int_0^E \frac{v(G)}{\sqrt{E-G}} dG = 0$$ since \ $v(G)$ \ is analytic. The other integral can be written 
$$\int_a^b \frac{u(G)\sqrt{2G}}{\sqrt{E-G(x)}} g(x) dx = \int_a^0 \frac{u(G)\sqrt{2G}}{\sqrt{E-G(x)}} g(x) dx + \int_0^b \frac{2u(G)\sqrt{2G}}{\sqrt{E-G(x)}} g(x) dx$$
{\footnotesize $$=\int_0^b \frac{u(G)\sqrt{2G(A(y))}}{\sqrt{E-G(y)}} g(A(y)) A'(y) dy = \int_0^b \frac{-u(G)\sqrt{2G(A(y))}}{\sqrt{E-G(y)}} g(y) dy = \int_0^b \frac{u(G)\sqrt{2G(y)}}{\sqrt{E-G(y)}} g(y) dy$$} since $\sqrt{2G(A(y))} = - \sqrt{2G(y)}.$
Finally, 
$$\int_a^b \frac{\phi(x)}{\sqrt{E-G(x)}} g(x) dx = \int_0^b \frac{u(G)\sqrt{2G}}{\sqrt{E-G(x)}} g(x) dx + \int_0^b \frac{u(G)\sqrt{2G}}{\sqrt{E-G(x)}} g(x) dx$$\\

\bigskip

{\bf Corollary 3-10}\quad {\it Under hypotheses of Proposition 3-7, consider the derivatives of $g$ : $g^{(j)}(x)=\frac{d^j g}{dx^j}$\ Then the analytic function  $$V_{m\nu}(x)= \prod_{j=1}^{m} \left(\frac{d^j g}{dx^j}\right)^\nu_j $$ may be expressed under the form :\\
$$V_{m,\nu}(x) = u_{m,\nu}(G)\sqrt{2G} + v_{m,\nu}(G)$$ where $\nu = (\nu_1,\nu_2,....,\nu_m)$ \ and \ $u_{m,\nu}$ and $v_{m,\nu}$ are analytic functions with respect to $G$.}
 
 \bigskip  

{\it Proof} \quad By Proposition 3-8 any derivative of $g$ may be written when $G$ is isochronous 
$$g^{(n)}(x) = a_{n}(G) \sqrt{2G} + b_{n}(G), n \geq 0$$
where \ $a_n $ and $b_n$ being analytic functions. It is easy to realize that it is the same for any power of any derivative $(g^{(n)}(x))^\nu_n$. We may prove that by recurrence $$(g^{(n)}(x))^\nu_n = a_{n,\nu}(G) \sqrt{2G} + b_{n,\nu}(G).$$ 
More generally, we may also prove by recurrence that a product of power of derivatives have the similar expression 
$$(g^{(n)}(x))^\nu_1 (g^{(p)}(x))^\nu_2 = a_{n,p,\nu}(G) \sqrt{2G} + b_{n,p,\nu}(G).$$
Thus we may write for any product $$V_{m,\nu}(x) = \prod_{j=1}^{m} \left(\frac{d^j g}{dx^j}\right)^\nu_j = u_{m,\nu}(G)\sqrt{2G} + v_{m,\nu}(G)$$ 

\section{The quantum spectrum qnd WKB quantization }

The semiclassical method of torus quantization is the first term of a certain $\hbar$-expansion, usually called the WKB expansion. This method due to Dunham is often  employed to study the spectra of quantum systems and to calculate the eigenvalues to any order. It is a generalisation of the Bohr-Sommerfeld quantisation condition.
 It is known, the semiclassical spectrum is perfectly regularly spaced. The semiclassical approximation allows us to understand the correspondence between quantum features and classical mechanics. We may expect that the limit $\hbar \rightarrow 0$ is equivalent to the limit $E \rightarrow \infty$ so that the error of the leading order semiclassical approximation should converge to zero as energy increases. \\ We need to know the first term of a certain $ h $-expansion, whose higher terms can be calculated with a recursion formula in one degree systems. We perform a systematic WKB expansion to all orders resulting in a convergent series whose sum is identical to the exact spectrum.\\
Two quantum systems with bounded classical analogs are said to be spectrally equivalent if their energy levels are identical. It is well known that the isoperiodicity equivalence is the classical version of the quantum isospectrality condition. In the same way it is obvious that the quantum counterpart of isochronicity is the harmonic spectrum (for regular potentials).  However, there exists many isochronous classical systems which do not have equally spaced quantum energy spectra.\\

In what follows, we briefly recall the WKB method before to make use of the first terms derived in [13] and from results of the previous Part 3 we will give an estimate of the n-th WKB correction.

\subsection{WBK series to all orders}
Following Robnik and Romanovski [14] consider the Schrodinger equation

$$\left[-\frac {\hbar^2}{2}\frac {d^2}{dx^2} + G(x)\right] \psi (x) = E \psi (x).$$
The Hamiltonian of the system is given by 
$$H = \frac{p^2}{m} + G(x)$$
where the mass $m=1$.\\
This Hamiltonian is a constant of motion, whose value is equal to the total energy E.\\
To calculate all the terms of the WKB expansion one observes that the wave function can always be written as 
$$\psi (x) = exp(\frac {i}{h}\sigma (x))$$
where the phase $\sigma(x)$ is a complex function verifying the Riccati differential equation 
$$\sigma'^2(x) + (\frac {i}{\hbar})\sigma''(x) = 2(E-G(x)).$$
The WKB expansion for the phase is a power series in $hbar$ :
$$\sigma(x) = \sum_0^\infty (\frac{\hbar}{i})^k \sigma_k (x).$$
After replacing and comparing like power of $hbar$  we obtain the recursion relation 
$${\sigma'_0}^2 = 2(E - G)\quad and \quad \sum_{k=0}^l \sigma'_k\sigma'_{n-k} + \sigma'_{n-1} =0, \ n \geq 1.$$ 
The quantization condition is obtained by requiring the uniqueness of the wave function
$$\int_\gamma d\sigma = \sum_{k=0}^\infty (\frac {i}{\hbar})^k \int_\gamma d\sigma_k = 2\pi \hbar n, \quad n \in N$$
where the integration contour $\gamma$ surrounds the two turning points of $G(x)$ at energy $E$.\\
The first term of this serie is the Bohr-Sommerfeld formula. It is proportional to the action $I_0(E)$
$$\int_\gamma d\sigma_0 = 2 \int_a^b \sqrt{2(E-G(x)}
dx = 2 \pi I_0(E).$$
The second term is the Maslov corrections. It can be shown to be equal to 
$$(\frac{\hbar}{i})\int_\gamma d\sigma_1 = - \pi \hbar.$$
All the other odd terms vanish. Because 
 these terms \ $\sigma'_{2k+1}, \ k \geq 1, $ are total derivatives and so \ $\int_\gamma \sigma'_{2k+1} = 0.$ This permits to rewrite the quantisation condition as
$$\sum_0^\infty I_{2k}(E) = (n+\frac{1}{2}) \hbar, \ n \in N$$
where $$ I_{2k}(E) = \frac {1}{2\pi}(\frac{\hbar}{i})^{2k} \int_\gamma d\sigma_{2k},\ k \in N.$$

When $G(x)$ is analytic and $x g(x) > 0$, it has been proved in [14] that the contour integrals can be replaced by equivalent Rieman integrals between the two turning points. More precisely, they proved the following
$$I_2(E) = -\frac{\hbar^2}{24\sqrt 2 \pi}\frac {\partial^2}{\partial E^2} \int_a^b \frac{g^2(x)}{\sqrt{E-G(x)}} dx$$ and 
$$I_4(E) = -\frac{\hbar^4}{4\sqrt 2 \pi}[\frac {1}{120}\frac {\partial^3}{\partial E^3} \int_a^b \frac{g'^2(x)}{\sqrt{E-G(x)}} dx $$ $$- \frac {1}{288}\frac {\partial^4}{\partial E^4} \int_a^b \frac{g^2(x) g'(x)}{\sqrt{E-G(x)}} dx]$$

Robnik and Salashnich [13] elaborated an efficient algorithm in order to simplify the function $d\sigma_k$ and produced the following formula for even $m$

$$\sigma'_m = \sum_{L(\nu)=m} \ \frac{2^{\frac{m}{2}-1+\mid\nu\mid}i}{(m-3+2\mid\nu\mid)!!} \frac{\partial^{\frac{m}{2}-1+\mid\nu\mid}}{\partial E^{\frac{m}{2}-1+\mid\nu\mid}} \frac{U_\nu G^{(\nu)}}{\sqrt{E-G}}$$
where $\nu = (\nu_1,\nu_2,...\nu_m), \nu_j \in N, \ L(\nu) = \sum_{j=1}^{2m} j v_j, \ and \ \mid\nu\mid =  \sum_{j=1}^{2m}  v_j$.\\
Moreover, they proved
$$\oint_\gamma d\sigma_m = 2 \sum_{L(\nu)=m} \ \frac{2^{\frac{m}{2}-1+\mid\nu\mid}i}{(m-3+2\mid\nu\mid)!!} \frac{\partial^{\frac{m}{2}-1+\mid\nu\mid}}{\partial E^{\frac{m}{2}-1+\mid\nu\mid}} \int_a^b \frac{U_\nu G^{(\nu)}}{\sqrt{E-G}} dx$$
where  $$G^{(\nu)}(x)= \prod_{j=1}^{m}  \left(\frac{d^j G}{dx^j}\right)^{\nu}_j$$
Coefficients $U_\nu$ are defined by a recurrence equation but no play a role in deriving the result.\\
Notice that higher order corrections quickly increase in complexity and we know only
few cases where a WKB expansion can be worked to all orders. Which resulting in a convergent series whose sum is identical to the exact
spectrum.

\subsection{WBK series for isochronous potentials}

It is  known, the spectrum of a potential is generally not strictly regularly spaced, in contrast to the harmonic one. The one-parameter family isotonic oscillator 
$$G(x) = \frac{1}{8 \alpha ^2} [\alpha x +1 - \frac{1}{\alpha x+ 1}]^2$$
is it an isochronous potential with a strictly equispaced (harmonic) spectrum.  Notice that as already proved in [13, 14], all the terms
$I_{2n}(E), n \geq 1 $ of the isotonic potential are constant and
once summed, the WKB series leads to the exact quantization
condition for this potential.  Moreover, the fact that the $I_{2n}(E), n \geq 1 $, are constant
ensures that the spectrum is strictly equispaced. The
interesting question is whether this family is the only one to be both classical and quantum "harmonic".  Dorignac [7] has provided a nice quantitative way to analyse the spectrum of analytic isochronous potentials.  This author descripted examples of isochronous potentials with equispaced spectrum Moreover, he proved scaling properties for any isochronous potential $V (x)$. indeed, there exists a simple scaling transformation, namely $\tilde V (\beta; x) = V (\beta x)/\beta^2$, which preserves its isochronism as well as its frequency.\\ 
We now use the preceding results of Part { 3} in order to express $I_2(E)$ and $I_4(E)$ as well as the nth correction $I_n(E)$ in terms related to the isochronous potential $G(x)$ and its derivatives.\\

By Proposition 3-8, we may write  \ $g=\frac{dG}{dx} = {a(G)\sqrt{2G}}+ b(G)$ . Writing 
{\footnotesize $$I_2(E) = \frac{-\hbar^2}{24\sqrt 2 \pi}\frac {\partial^2}{\partial E^2} \int_a^b \frac{g^2(x)}{\sqrt{E-G(x)}} dx =  \frac{-\hbar^2}{24\sqrt 2 \pi}\frac {\partial^2}{\partial E^2} \int_a^b \frac{g(x)}{\sqrt{E-G(x)}} g(x)dx$$} and by Proposition 3-9 we may express
$$ -\frac{\hbar^2}{24\sqrt 2 \pi}\frac {\partial^2}{\partial E^2} \int_0^b \frac{2a(G)\sqrt{2G}}{\sqrt{E-G(x)}}g(x) dx =  -\frac{\hbar^2}{24\sqrt 2 \pi}\frac {\partial^2}{\partial E^2} \int_0^E \frac{2a(v)\sqrt{2v}}{\sqrt{E-v}}dv.$$
 Then making the change of variables $ u=\frac{v}{E}$ (we suppose here $\omega = 1$)
$$I_2(E) =   -\frac{\hbar^2}{24 \pi}\frac {\partial^2}{\partial E^2} [ E\int_0^1 \frac{2a(u E)\sqrt{u}}{\sqrt{1-u}}\ du].$$
Another formulation is given by an Abel type integral
$$I_2(E) =   -\frac{\hbar^2}{24 \pi} \frac{1}{E^2} \int_0^E \frac{({\sqrt{2v}})^3}{\sqrt{E-v}}\ \frac{d^2}{dv^2} [v a(v)] $$ $$= -\frac{\hbar^2}{24 \pi} \frac{1}{E^2} \int_0^E \frac{({\sqrt{2v}})^3}{\sqrt{E-v}}\ [2\frac{da}{dG}(v)+ v \frac{d^2a}{dG^2}(v)].$$

A similar calculation gives the fourth order correction (see Appendix A for details )
$$I_4(E) = -\frac{\hbar^4}{4\sqrt 2 \pi}[\frac {1}{120}\frac {\partial^3}{\partial E^3} \int_a^b \frac{g'^2(x)}{g(x)\sqrt{E-G(x)}} g(x) dx $$ $$- \frac {1}{288}\frac {\partial^4}{\partial E^4} \int_a^b \frac{g(x) g'(x)}{\sqrt{E-G(x)}} g(x) dx].$$
By proposition 3-8 one gets $g=\frac{dG}{dx} = {a(G)\sqrt{2G}}+ b(G)$\\ and $g'=\frac{dg}{dx} = {a_1(G)\sqrt{2G}}+ b_1(G)$. However, we may easily prove  
$$\frac{g'^2}{g} = a_{1,2}(G)\sqrt{2G}+ b_{1,2}(G)$$ where $a_{1,2}(G)$ and $b_{1,2}(G)$ are analytic functions. As well as 
$$g(x) g'(x) = c_{1,2}(G)\sqrt{2G}+ d_{1,2}(G)$$ where $c_{1,2}(G)$ and $c_{1,2}(G)$ are analytic functions.
Similar to $I_2, I_4$ may be expressed through Abel integrals :

$$I_4(E) = -\frac{\hbar^4}{4 \pi}[\frac {E^{-3}}{120} \int_0^E \frac{(\sqrt{2v})^5}{\sqrt{E-v}} \frac {\partial^3}{\partial v^3} a_{1,2}(v) \ dv $$ $$- \frac {E^{-4}}{288} \int_0^E \frac{(\sqrt{2v})^7}{\sqrt{E-v}} \frac {\partial^4}{\partial v^4} c_{1,2}(v)\ dv].$$

Starting from expression  for $I_2(E)$ and using the properties of Abel transforms, as remarked by [7] its possible to invert the problem and to calculate the
general expression of the functions $a_p(G)$ and $b_p(G)$  corresponding to a prescribed function $I_2(E)$. One can choose $I_2(E)$ (and deduce
the corresponding analytic isochronous potential) such that, its asymptotic decay is faster than the asymptotic decay of $I_4(E)$. Therefore, $I_2(E)$ and $I_4(E)$ grow
exponentially fast as $E$ grows to $\infty$.\\ As we will show below these conclusions are similar to higher order corrections.\\

Turn now to upper order WKB correction. The explicit expression for $I_{2n}(E)$ is given by
$$I_{2n}(E)= -\frac{\sqrt 2}{\pi} \hbar^{2n} \sum_{L(\nu)=2n} \ \frac{2^{\mid\nu\mid}}{(2n-3+2\mid\nu\mid)!!} J_\nu (E)$$
where 
$$J_\nu (E) = \frac{\partial^{n-1+\mid\nu\mid}}{\partial E^{n-1+\mid\nu\mid}} \int_a^b \frac{U_\nu G^{(\nu)}}{\sqrt{E-G}} dx$$
where  $$G^{(\nu)}(x)= \prod_{j=1}^{2n}  (\frac{d^j G}{dx^j})^{\nu}_j$$
and where $\nu = (\nu_1,\nu_2,...,\nu_{2n}), \nu_j \in N, L(\nu)=\sum_{j=1}^{2n} j\nu_j$\ and \ $\mid\nu\mid = \sum_{j=1}^{2n} \nu_j$. The coefficients $U_\nu$ satisfy a certain recurrence relation not useful for the sequel\\

By Corollary 3-10 \ $G^{(\nu)}$\ may be expressed under the form 
$$G^{(\nu)}(x)= \prod_{j=1}^{m}  \left(\frac{d^j G}{dx^j}\right)^{\nu}_j = u_{n,\nu}(G)\sqrt{2G} + v_{n,\nu}(G)$$
where $u_{n,\nu}$ and $v_{n,\nu}$ are analytic functions with respect to $G$.
Therefore,
$$J_\nu (E) = \frac{\partial^{n-1+\mid\nu\mid}}{\partial E^{n-1+\mid\nu\mid}} \int_a^b \frac{U_\nu u_{n,\nu}(G)\sqrt{2G}}{\sqrt{E-G}} g(x) dx$$
By Corollary 3-9, we can write
$$J_\nu (E) = \frac{\partial^{n-1+\mid\nu\mid}}{\partial E^{n-1+\mid\nu\mid}} \int_0^b \frac{2 U_\nu u_{n,\nu}(G)\sqrt{2G}}{\sqrt{E-G}}g(x) dx$$
$$J_\nu (E) = \frac{\partial^{n-1+\mid\nu\mid}}{\partial E^{n-1+\mid\nu\mid}} \int_0^E \frac{2 U_\nu u_{n,\nu}(v)\sqrt{2v}}{\sqrt{E-v}}dv$$
Another equivalent formulation via Abel integrals
$$J_\nu (E) = 2 U_\nu E^{{-n+1-\mid\nu\mid}} \int_0^E \frac{(\sqrt{2v})^{n-2+\mid\nu\mid}}{\sqrt{E-v}}\frac{\partial^{n-1+\mid\nu\mid}}{\partial E^{n-1+\mid\nu\mid}} u_{n,\nu}(v) dv$$

Similar to $I_2$ and $ I_4$\ the nth correction\ $ I_{2n}$  seems too to be expressed through Abel integrals :

$${\footnotesize I_{2n}(E)= \frac{-\hbar^{2n}}{\pi}  \sum_{L(\nu)=2n} \ \frac{2^{\mid\nu\mid +1} U_\nu E^{{-n+1-\mid\nu\mid}}}{(2n-3+2\mid\nu\mid)!!}   \int_0^E \frac{(\sqrt{2v})^{n-2+\mid\nu\mid}}{\sqrt{E-v}}\frac{\partial^{n-1+\mid\nu\mid}}{\partial E^{n-1+\mid\nu\mid}} u_{n,\nu}(v) dv}$$

Thus, the WKB corrections $I_{2n}(E)$ grow exponentially fast as $E$ grows to $\infty$.
 The calculation above suggest the entire WKB series could be summed for any isochronous potential and would 
be finite as $E$ grows to $\infty$.
 
\bigskip

\section{On the parametrization of isochronous potentials}
The purpose of this section is to offer some refinements of results reported in [5] and to highlight natural conditions of isochronism.\\
Let us write $$G(x) =  \frac{1}{2}x^2+ G_1(x)+G_2(x)$$ where the function \ $G_1(x)= \sum_{k\geq 2} \frac{a_{2k-2}}{2k-1}x^{2k-1}$ \ is odd and \\ $G_2(x) = \sum_{k\geq 2} \frac{a_{2k-1}}{2k}x^{2k}$\ is even.\\ 
Let us write
$$g(x)= x+\sum_{n\geq 2}a_n x^n\qquad and \qquad  G(x) = \frac{1}{2}x^2 + \sum_{n\geq 2} \frac {a_n}{n+1} x^{n+1}.$$  In [5]  we proved the following

\bigskip

{\bf Theorem C}\quad {\it Let the analytic potential $$G(x) = \frac{1}{2}x^2+\sum_{n\geq 3} \frac{a_{n-1}}{n}x^n=\frac{1}{2}x^2+G_1(x)+G_2(x)$$ of equation $$(1)\qquad \ddot x+g(x)=0 $$ where $G$ is defined in a neighborhood of $0$ and $x g(x) > 0 , \ for\ x\neq 0$.\\
Suppose $G(x)$ is isochronous then the odd coefficients of the expansion of $g(x)$ can be expressed in terms of rational polynomials involving the even coefficients:  
$$ a_{2k+1}= f(a_{2k},a_{2k-2},...,a_2).  $$ In particular,  $G_2(x)\equiv 0$ is equivalent to $G_1(x)\equiv 0$, i.e. $G(x)= \frac{1}{2} x^2$ is harmonic.}\\

\bigskip

This means that once the odd part $G_1$ of an analytic isochronous potential is specified then the even part $G_2$ is completely determined.

In this context as we have seen foregoing  the successive derivatives of $G(x)$ can be written \ $g^{(n)}(x) = a_{n}(G) \sqrt{2G} + b_{n}(G), n \geq 0.$\ This  means that the analytic functions may be only depending on the odd part $G_1$ \  $a_{n}(G)=a_{n}(G_1) $\ and \  $b_{n}(G)=b_{n}(G_1). $\ This can significantly simplify the problem. this remark is also valid for all the powers of derivatives products 
$$G^{(\nu)}(x)= \prod_{j=1}^{2m}  \left(\frac{d^j G}{dx^j}\right)^{\nu}_j = u_{m,\nu}(G_1)\sqrt{2G} + v_{m,\nu}(G_1)$$ where $\nu = (\nu_1,\nu_2,....,\nu_m)$ \ and \ $u_{m,\nu}$ and $v_{m,\nu}$ are analytic functions with respect to $G_1$.

The first terms of the expansion are:

$$a_3=\frac{10}{9}a_2^2,\qquad a_5=\frac{14}{5}a_2a_4-\frac{56}{27}a_2^4,$$ 

$$ a_7=\frac{ -592}{45}a_4a_2^3+\frac{848}{81}a_2^6+\frac{24}{7}a_2a_6+\frac{36}{25}a_4^2,$$

\bigskip

{\bf Remarks 5-1}\\
Consider now condition (7) then equality 
$$(10)\qquad x - \frac{2 G}{g} = F(G) = b_0 + b_1 G+ b_2 G^2+ b_3 G^3+ b_4 G^4+ ...$$ 
(where $F(G)$ is analytic) ensures the isochronicity of the potential $G$. \\ 
It is also possible to express coefficients $a_n$ of the expansion of $g(x) = \frac {dG(x)}{dx}$  in terms of $b_0, b_1, b_3,..$ \\
Indeed, after replacing expressions of $g$ and $G$ in  $$x g(x) - 2G(x) = g(x) F(G) $$ and by identifying
$$ x^2+\sum_{n\geq 2}a_n x^{n+1} - 2G  = [ x+\sum_{n\geq 2}a_n x^n ][b_0 + b_1 G+ b_2 G^2+ b_3 G^3+ b_4 G^4+ ...] .$$

We then find
$$G(x) = \frac{1}{2}\,{x}^{2}-\frac{1}{2}\,b_{0}{x}^{3}+\frac{5}{8}\,{b_{0}}^{2}{x}^{4}-
 \left({\frac {1}{24}}\,b_{1}+{\frac {7}{8}}\,{b_{0}}^{3}
 \right) {x}^{5}+{\frac {7b_{{0}}}{270}} \left( {\frac {405}{8}}\,{b
_{0}}^{3}+{\frac {45}{8}}\,b_{1} \right) {x}^{6}+...$$
In fact, only the odd part plays a role
$$G_1(x) = -\frac{1}{2}\,b_{0}{x}^{3}+
 \left(-{\frac {1}{24}}\,b_{1}-{\frac {7}{8}}\,{b_{0}}^{3}
 \right) {x}^{5}+...$$ \\

\section{Applications to family of isochronous potentials}

It is well known that the harmonic potential \ $ G=\frac{1}{2} x^2$\ is the only polynomial potential which is isochronous. Concerning the rational case, Chalykh and Veselov [3] proved that a rational potential $G(x)$ (which is not a polynomial) is isochronous if and only if 
$$G(x) = \frac{1}{8 \alpha ^2} [\alpha x +1 - \frac{1}{\alpha x+ 1}]^2.$$

\subsection{ A three-parameters family of isochronous \\ potentials}

We now apply the above considerations to the determination
of isochronous potentials whose analytical expression
can be given explicitly. We will use for that Theorem A or B and their corollaries. \\

To be concrete, we derive at first a three-parameters family of potentials which have described by another way in [5].  \\

Let us consider 

$$\frac{2G}{g} - x = F(G) =2\,{\frac {a \left( -1+\sqrt {1+b G} \right) }{b \sqrt {1+b G}}} ,$$ where $a $ and $b $ are real parameters such that \ $2a^2 \leq b.$\\
One have seen that $F$ and $P$ are relied by $F(G) = 2G\frac{dP(G)}{dG} - P(G).$ 
We may deduce the analytic function $P$ 
$$P(G) = {\frac {-2\,a+\sqrt {4\,{a}^{2}+4\,G b{a}^{2}}}{b}}. $$ 
Thus 
$$x= \sqrt{2G} + P(G) = {\frac {-2\,a+\sqrt{2G}b+\sqrt {4\,{a}^{2}+4\,G b{a}^{2}}}{b}} $$
 A resolution of these equations yields
\begin{equation}G(x) =\frac {8a^2+(b+2a^2)(4a x+b x^2)-(4a^2+2ab x)\sqrt{2(2+b x^2+4a x)}}{2(b-2a^2)^2}   \end{equation}
Then, the above potential is isochronous according to Theorem B. 
 Applying scaling property of isochronous potentials. The potentials $G(x)$ and $\frac {1}{c^2}G(c x)$ have the same period. That means the following three-parameters potentials family is isochronous 
\begin{equation} G(x) = \frac {1}{2c^2}X^2(c x) = \frac{[2a+b c x- a \sqrt{2(2+b c^2 x^2+4a c x)}]^2}{2c^2(b-2a^2)^2}\end{equation}
  
Many special cases of this 3-parameters family have been described in [5]. Moreover, the associated involution  is
$$A(x) =  \frac{2a^2x - 2a-b c x+ a \sqrt{2(2+b c^2 x^2+4a c x)}}{c(b-2a^2)}$$

\bigskip
\subsection{ Others isochronous potentials}

We will give here one-parameters families of potentials with constant period. Using scaling property in multiplying $x$ by a non zero real $\gamma$ : the potentials 
$G(x) $ and $\frac{1}{\gamma^2} G(\gamma x)$ have the same period. To our
knowledge, some families of potentials derived in this subsection
 never before appeared in the literature. However, we could not highlight families with more than one parameter. It would be interesting to find a family with at least 2-parameters including all the examples below.

\bigskip
{\bf 1 -} Let us consider the following example $$\frac{2G}{g} - x = F(G) =  -\frac{c (6G+1)}{(1+2G)^2}$$
where $c \neq 0$ is a parameter. Since one have seen $F(G) = 2G\frac{dP(G)}{dG} - P(G)$ it yields
$$x= \sqrt {2G} + P(G) = \sqrt {2G} - \frac{c }{1+2G}$$
and $$\frac{1}{g} = \frac{1}{\sqrt {2G}}+ P'(G) =  \frac{1}{\sqrt {2G}}+ \frac{2c}{(1+2G)^2}.$$
Solving the equation \ $x= \sqrt {2G} - \frac{c }{1+2G}$\  and writing 

$${\footnotesize u(x)=\,\sqrt [3]{18\,\sqrt {2}x+27\,\sqrt {2}c+2\,\sqrt {2}{x}
^{3}+3\,\sqrt {24+48\,{x}^{2}+24\,{x}^{4}+216\,xc+162\,{c}
^{2}+24\,c{x}^{3}} }}$$
Thanks to {\it Maple} we find the potential which is isochronous by Theorem B
$$ G= \left( \frac{1}{6}u(x)-\,{\frac {1-\frac{1}{3}\,{x}^{2}}{u(x)}}+\,\frac{\sqrt {2}x}{6} \right) ^{2}.$$
Its involution $A$ is 
$$A(x) = \frac{x}{3} - 2 \sqrt{2} \left( \frac{1}{3}u(x)-\,{\frac {6-2\,{x}^{2}}{3u(x)}} \right).$$

\bigskip

{\bf 2 -} \ Let $$\frac{2G}{g} - x = F(G) = -2\,{\frac {-2a G }{ \left( 1+2\,{a}^{2}G
 \right) ^{3/2}}}$$
where the parameter $a\neq 0$.
The function $P'$  defined by\  
$\frac{1}{g} = \frac{1}{\sqrt {2G}}+ P'(G) $\  may be written
$$P'(G) = 2\,{\frac {a \left( 1+{a}^{2}G \right) }{ \left( 1+2\,{a}^{2}G
 \right) ^{3/2}}}.$$
Its integral which verifies \ $x = \sqrt {2G} + P(G)$\ has the following form
$$P(G) =2\,{\frac {aG}{\sqrt {1+2\,{a}^{2}G}}}.$$
If we write
{\footnotesize $$ u(x)=\sqrt [3]{-15\,{a}^{2}{x}^{2}+39\,{a}^{4}{x}^{4}+1+{a}^{6}{x}^{6}+6\,
\sqrt {3}\sqrt {-1+11\,{a}^{2}{x}^{2}+{a}^{4}{x}^{4}}{a}^{3}{x}^{3}}$$}

Then we find the isochronous potential
$$ G= \frac{1}{2}[\,{\frac {u(x)}{6{a}^{2}x}}+\,{\frac {-10\,{a}^{2}{x}^{2}+1+{a}^{4}{x}
^{4}}{6{a}^{2}x\sqrt [3]{u(x)}}}+\,{\frac {1+{a}^{2}{x}^{2}}{6{a}^{2}x}}]^2.$$
The involution associated to this potential is
$$A(x) = -\,{\frac {u(x)}{3{a}^{2}x}}-\,{\frac {-10\,{a}^{2}{x}^{2}+1+{a}^{4}{x}
^{4}}{3{a}^{2}x\sqrt [3]{u(x)}}}+\,{\frac {-1+2{a}^{2}{x}^{2}}{3{a}^{2}x}}$$

\bigskip

{\bf 3 -} \  Let $$\frac{2G}{g} - x = F(G) = {-\frac {a-2\,G+4\,Ga}{ \left( 1+2\,G \right) ^{3/2}}}$$
where the parameter $a\neq 0$. By $F(G) = 2G\frac{dP(G)}{dG} - P(G)$ one gets the function $P'$ 
$$P'(G) ={\frac {2+2\,G-a}{ \left( 1+2\,G \right) ^{3/2}}}.$$
Its integral $P$ is
$$P(G)={\frac {a+2\,G}{\sqrt {1+2\,G}}}-a$$
Let us denote
 $$v(x)=-6\,
{x}^{6}a+6\,{x}^{4}{a}^{3}-{x}^{6}{a}^{2}-{a}^{2}+{x}^{8}-12\,{a}^{4}+
6\,{a}^{3}+23\,{x}^{2}{a}^{2}-2\,{x}^{2}a-$$ $${x}^{4}+11\,{x}^{6}+15\,{a}^
{4}{x}^{2}+8\,{a}^{5}-32\,{a}^{3}{x}^{2}-27\,{a}^{2}{x}^{4}+28\,a{x}^{
4}$$
 {\footnotesize $$u(x)=[-15\,{x}^{2}+39\,{x}^{4}+24\,{x}^{2}a-42\,{x}^{2}{a}^{2}+1-6
\,a+12\,{a}^{2}+{x}^{6}-6\,a{x}^{4}-8\,{a}^{3}+6\,\sqrt {3}\sqrt {v(x)}x]^3$$}
Then the associate isochronous potential is
$$G(x)=\frac{1}{2} [\,{\frac {u(x)}{6x}}+\,{\frac {-10\,{x}^{2}+1-4\,a+{x}^{4}-4\,{x}^{2
}a+4\,{a}^{2}}{6xu(x)}}-\,{\frac {-1-{x}^{2}+2\,a}{6x}}]^2.$$
Moreover,
$$A(x) = -\,{\frac {u(x)}{3x}}-\,{\frac {-10\,{x}^{2}+1-4\,a+{x}^{4}-4\,{x}^{2
}a+4\,{a}^{2}}{3xu(x)}}+\,{\frac {-1+2{x}^{2}+2\,a}{3x}}.$$

\bigskip
{\bf 4 -}  
Let us consider  
$$ \frac{2G}{g} - x = F(G) = {\frac {\alpha}{ \left( 1+2 \beta^2 G \right) ^{3/2}}}.$$

Then the function $P'$ which verifies \  $\frac{1}{g} = \frac{1}{\sqrt {2G}}+ P'(G) 
$\ may be written

$$P'(G)=\,{\frac { \left( 3+4\,{b}^{2}G \right) a\,{b}^{2}}{ \left( 1+2\,{b}^{2}G
 \right) ^{3/2}}}.
$$

 Its integral is
$$P(G)=\,{\frac {\alpha \left( 1+2\,(2G){3\beta}^{2} \right) }{{\sqrt {1+(2G){\beta}^{2}}}} - {\alpha}}$$

Taking for example\ $\alpha={1}{2}$\ and thanks to {\it Maple} we find another isochronous potential. \\

Let the following functions\quad 
{\footnotesize $$v(x)=\sqrt {1024\,{x
}^{6}{\alpha}^{6}+4608\,{x}^{5}{\alpha}^{5}+26496\,{x}^{4}{\alpha}^{4}+62208\,{x}^{3}
{\alpha}^{3}+101412\,{x}^{2}{\alpha}^{2}+86022\,x\alpha+45927}$$}
   {\small $$ u(x)= -5751+31536\,x\alpha+21600\,{x}^{2}{\alpha}^{2
}+768\,{x}^{3}{\alpha}^{3}+256\,{x}^{4}{\alpha}^{4}+72\,\sqrt {3} v(x) $$}

Solving $x =\sqrt {2G}+P(G) $  and in using scaling property the isochronous potential is

 {\small $$G(x,\gamma) = (\frac{1}{2}) [{\frac { [u(\gamma x)]^{1/3}}{24\alpha \left( 3+4\,\gamma x \alpha \right) }}+{\frac { \left( 16
\,{\gamma^2 x}^{2}{\alpha}^{2}+24\,\gamma x \alpha-423 \right)  \left( 3+4\,\gamma x \alpha \right) }{24\alpha [u(\gamma x)]^{1/3}}+
{\frac {3+4\,\gamma x \alpha}{24\alpha}}}]^2.$$}
Its involution is (for $\gamma =1$)
$$A(x) = -{\frac { [u( x)]^{1/3}}{12\alpha \left( 3+4\, x \alpha \right) }}-{\frac { \left( 16
\,{ x}^{2}{\alpha}^{2}+24\, x \alpha-423 \right)  \left(3+4\, x \alpha \right) }{12\alpha [u( x)]^{1/3}}+
{\frac { 8\, x \alpha -3}{12\alpha}}}.$$ 

\newpage

{\bf \Large Appendix }

\bigskip
The purpose of this appendix is to show that the first 2 corrections WKB can be evaluated in a natural way thanks to Corollary 3-10. Which gives another, more direct approach. This same method can also be applied to higher order corrections $I_n$ since we have seen that the integrals are also of Abel type.\\

The fourth WKB correction $I_4$ is 
$$I_4(E) = -\frac{\hbar^4}{4\sqrt 2 \pi}[\frac {1}{120}\frac {\partial^3}{\partial E^3} \int_a^b \frac{g'^2(x)}{g(x)\sqrt{E-G(x)}} g(x) dx $$ $$- \frac {1}{288}\frac {\partial^4}{\partial E^4} \int_a^b \frac{g(x) g'(x)}{\sqrt{E-G(x)}} g(x) dx]$$
 It turn out that an exact expression may be available. We shall justify the expressions of
$$\frac{g'^2}{g} = a_{1,2}(G)\sqrt{2G}+ b_{1,2}(G), \quad g(x) g'(x) = c_{1,2}(G)\sqrt{2G}+ d_{1,2}(G)$$ where $a_{1,2}(G),\ b_{1,2}(G),\ c_{1,2}(G)$ and $c_{1,2}(G)$ are analytic functions. We have seen that the isochronicity conditions are 
$$x = \sqrt{2G} + P(G)\quad  and \ (or) \quad \frac{1}{\sqrt{2G}} + P'(G) = \frac{1}{g}$$
It implies 
$$g' =\frac{d g}{dx} =2\,{\frac {G \left( \sqrt {2}\sqrt {G}+\phi \left( G \right) -2\,Gf
 \left( G \right)  \right) }{ \left( \sqrt {2}\sqrt {G}+\phi \left( G
 \right)  \right) ^{3}}}$$
where $\phi = 2G P'(G) $ and $f = P'(G)+2G P''(G).$ Let $A$ be the involution defined above by $G(A(x))= G(x) $. Recall that \ $\sqrt{2G(A(x))}=-\sqrt{2G(x)}.$\ Thus  {\footnotesize $$g g'(x) - g g'(A(x)) = 4\,{\frac {{G}^{2} \left( -\sqrt {2}\sqrt {G}+\phi \left( G \right) -2
\,Gf \left( G \right)  \right) }{ \left( -\sqrt {2}\sqrt {G}+\phi
 \left( G \right)  \right) ^{4}}}-4\,{\frac {{G}^{2} \left( \sqrt {2}
\sqrt {G}+\phi \left( G \right) -2\,Gf \left( G \right)  \right) }{
 \left( \sqrt {2}\sqrt {G}+\phi \left( G \right)  \right) ^{4}}}$$}
After simplification one has 
$$g g'(x) - g g'(A(x)) =$$ {\footnotesize $$8\,{\frac {{G}^{5/2}\sqrt {2} \left( -4\,{G}^{2}-4\,G \left( \phi
 \left( G \right)  \right) ^{2}+3\, \left( \phi \left( G \right) 
 \right) ^{4}-16\,{G}^{2}f \left( G \right) \phi \left( G \right) -8\,
 \left( \phi \left( G \right)  \right) ^{3}Gf \left( G \right) 
 \right) }{ \left( \sqrt {2}\sqrt {G}-\phi \left( G \right)  \right) ^
{4} \left( \sqrt {2}\sqrt {G}+\phi \left( G \right)  \right) ^{4}}}= 2 c_{1,2}(G)\sqrt{2G}$$}
By the same way one obtains 
$$ g g'(x) + g g'(A(x))=$$ {\footnotesize $$ \,{\frac {8{G}^{2} \left( -12\,{G}^{2}\phi \left( G \right) +4\,G
 \left( \phi \left( G \right)  \right) ^{3}+ \left( \phi \left( G
 \right)  \right) ^{5}-8\,{G}^{3}f \left( G \right) -24\,{G}^{2}f
 \left( G \right)  \left( \phi \left( G \right)  \right) ^{2}-2\,Gf
 \left( G \right)  \left( \phi \left( G \right)  \right) ^{4} \right) 
}{ \left( \sqrt {2}\sqrt {G}-\phi \left( G \right)  \right) ^{4}
 \left( \sqrt {2}\sqrt {G}+\phi \left( G \right)  \right) ^{4}}}$$}
$$=2 d_{1,2}(G)= \frac{64G^5(-3P'+4GP'^3+4G^2P'5-G^3f-12GfP'2-4G^2fP'4)}{ \left( \sqrt {2}\sqrt {G}-\phi \left( G \right)  \right) ^{4}
 \left( \sqrt {2}\sqrt {G}+\phi \left( G \right)  \right) ^{4}}$$

Turn now to the other expression. A calculation yields
$$\frac{g'^2}{g}=2\,{\frac {G \left( \sqrt {2}\sqrt {G}+\phi \left( G \right) -2\,Gf
 \left( G \right)  \right) ^{2}}{ \left( \sqrt {2}\sqrt {G}+\phi
 \left( G \right)  \right) ^{5}}}$$

{\footnotesize $$\frac{g'^2}{g}(x) - \frac{g'^2}{g}(A(x)) =2 a_{1,2}(G)\sqrt{2G} = $$  $$ -128\,{\frac {{G}^{9/2}\sqrt {2} \left( -{\it P'}-2\,G{{\it P'}}^{3}+G
{\it P''}+12\,{G}^{2}{{\it P'}}^{2}{\it P''}+4\,{G}^{3}{\it P''}\,{{\it 
P'}}^{4}-8\,{G}^{3}{\it P'}\,{{\it P''}}^{2}-16\,{G}^{4}{{\it P''}}^{2}{
{\it P'}}^{3} \right) }{ \left( \sqrt {2}\sqrt {G}-2\,G{\it P'}
 \right) ^{4} \left( \sqrt {2}\sqrt {G}+2\,G{\it P'} \right) ^{4}}}
$$ }
By the same way one obtains 

$$\frac{g'^2}{g}(x) + \frac{g'^2}{g}(A(x))=2 b_{1,2}(G) {\left( \sqrt {2}\sqrt {G}-2\,G{\it P'} \right) ^{-5}
 \left( \sqrt {2}\sqrt {G}+2\,G{\it P'} \right) ^{-5}} = $$ $$ 64\, G^5(-4\,G{\it P''}+4\,{G}^{2}{{\it P'}}^{5}+20
\,G{{\it P"}}^{3}+5\,{\it P'}-$$ $$80\,{G}^{2}{\it P''}\,{{\it P'}}^{2}-80\,
{G}^{3}{{\it P'}}^{4}{\it P''}+160\,{G}^{4}{{\it P'}}^{3}{{\it P''}}^{2}
+40\,{G}^{3}{{\it P''}}^{2}{\it P'}+32\,{G}^{4}{{\it P''}}^{2}{{\it P'}}
^{5} ) $$ 

Notice that the denominator is a power of \ ${\left( \sqrt {2}\sqrt {G}-2\,G{\it P'} \right) 
 \left( \sqrt {2}\sqrt {G}+2\,G{\it P'} \right) }$ which is equal to \ $2G - 4G^2 P'^2= 2G (1 - 2G P'^2).$
We may simplify by $G$ and $1 - 2G P'^2(G)$ is always non zero. This means that all the functions, \ $a_{1,2}(G), b_{1,2}(G), c_{1,2}(G)$ and $d_{1,2}(G)$ are analytic.\\

 \bigskip

{\bf References}\\                  
                     
[1]\ F. Calogero  \ {\it Isochronous systems}\  Oxford University Press, Oxford, (2008).  

\smallskip 

[2] Bolotin S and MacKay RS 2003,\quad  {\it Isochronous potentials}, in: "Localization and energy transfer in nonlinear systems",
eds L Vazquez, RS MacKay, M-P Zorzano, World, Sci., 217.

[3] \ O. Chalykh and A. Veselov \quad {\it  A remark on rational isochronous potentials} \quad J. Nonlinear Math. Phys. {\bf 12-1}, p. 179-183, (2005). 

\smallskip 

[4] Chavarriga J and Sabatini M 1999,\quad  {\it A survey of isochronous centers}, Qual. Theo. Dyn. Syst., 1, 1.

\smallskip

[5]\  R. Chouikha \quad {\it Period function and characterizations of Isochronous potentials} \quad  arXiv:1109.4611 (2011)
                        
\smallskip 

[6]  \ S.N. Chow and D. Wang \quad {\it On the monotonicity of the period function of some second order equations}\quad Casopis Pest. Mat. {\bf 111}, p. 14-25, (1986).

\smallskip

[7] \ J. Dorignac \quad {\it On the quantum spectrum of isochronous potentials}\quad J. Phys; A: Math. Gen., {\bf 38}, p. 6183-6210 (2005).

\smallskip

[8] Eleonskii VM, Korolev VG and Kulagin NE 1997,\quad  {\it On a classical analog of the isospectral Schrodinger problem},
JETP Lett., 65 (11), 889.

\smallskip

[9] Erd´elyi A, Magnus W, Oberhettinger F, Tricomi FG 1954,\quad  {\it Tables of integral transforms}, Vol. II, (New York,
McGraw-Hill book company, Inc.).

\smallskip

[10] \ I. Koukles and N. Piskounov \quad {\it Sur les vibrations tautochrones dans les systèmes conservatifs et non conservatifs}. C. R. Acad. Sci., URSS, vol XVII, n°9, p. 417-475, (1937).

\smallskip

[11]\ M. Kuczma \quad {\it Functional equations in a single variable} \quad Monografie Matematyczne, Tom 46, Warsaw, (1968).

\smallskip

[12]\ L.D. Landau E.M. Lifschitz \quad {\it Mechanics, Course of Theorical Physics} Vol 1, Pergamon Press, Oxford, (1960).                     

\smallskip

[13] Robnik M and Salasnich L 1997,\quad  {\it WKB to all orders and the accuracy of the semiclassical quantization}, J. Phys.
A: Math. Gen., 30, 1711.

\smallskip

[14] Robnik M and Romanovski VG 2000, \quad {\it Some properties of WKB series}, J. Phys. A: Math. Gen., 33, 5093.

\smallskip

[15] \ F. Rothe \quad {\it Remarks on periods of planar Hamiltonian systems.} \quad SIAM J. Math. Anal., {\bf 24}, p.129-154, (1993).

\smallskip
[16]\ FH. Stillinger and DK. Stillinger \quad {\it Pseudoharmonic oscillators and inadequacy of semiclassical quantization}\quad J. Phys. Chim., {\bf 93}, 6890, (1989).

\smallskip

[17]\  M. Urabe \quad  {\it The potential force yielding a periodic motion whose period is an arbitrary continuous function of the amplitude of the velocity}\ Arch. Ration. Mech. Anal.,{\bf 11}, p.27-33, (1962). 

\end{document}